\def\be{\begin{equation}}
\def\ee{\end{equation}}
\def\bea{\begin{eqnarray}}
\def\eea{\end{eqnarray}}
\def\bse{\begin{subequations}}
\def\ese{\end{subequations}}

\documentclass[prb,twocolumn,showpacs,amsmath,amssymb,eqsecnum]{revtex4}
\usepackage{graphicx}
\usepackage{dcolumn}
\usepackage{bm}
\begin{document}
\title{Fluctuation-induced first-order transition in $p\,$-wave superconductors
}
\author{Qi Li$^{1,2}$, D. Belitz$^2$, and John Toner$^2$}
\affiliation{$^1$ Department of Physics and Condensed Matter Theory Center,\\
                  University of Maryland, College Park, MD 20742\\
             $^2$ Department of Physics and
                  Institute of Theoretical Science,\\
                  University of Oregon, Eugene, OR 97403
         }

\date{\today}

\begin{abstract}
The problem of a fluctuation-induced first-order transition is considered for
$p\,$-wave superconductors. Both an $\epsilon$-expansion about $d=4$ and a
large-$n$ expansion conclude that the transition for the physical case $n=6$ in
$d=3$ is of first order, as in the $s$-wave case.
\end{abstract}

\pacs{64.60.ae,64.60.De,74.20.De }

\maketitle

\section{Introduction}
\label{sec:I}

BCS theory predicts that the phase transition from the normal state
to the superconducting state in $s$-wave superconductors is
continuous or second order. However, in 1974 Halperin, Lubensky, and
Ma \cite{Halperin_Lubensky_Ma_1974} argued, based on a $4-\epsilon$
expansion, that the coupling between the superconducting order
parameter and the electromagnetic vector potential drives the
transition first order. This conclusion is inevitable for extreme
type-I superconductors where fluctuations of the order parameter are
negligible and the vector potential can be integrated out exactly.
The mechanism in this case is known as a fluctuation-induced first
order transition, and it is analogous to the spontaneous mass
generation known in particle physics as the Coleman-Weinberg
mechanism.\cite{Coleman_Weinberg_1973} When order parameter
fluctuations cannot be neglected, and especially for type-II
superconductors, the problem cannot be solved exactly. The authors
of Ref.\ \onlinecite{Halperin_Lubensky_Ma_1974} generalized the
problem by considering an $n/2$-dimensional complex order parameter
and conducting a renormalization-group (RG) analysis in
$d=4-\epsilon$ dimensions. The physical case of interest is $n=2$
and $\epsilon=1$. To first order in $\epsilon$ they found that a RG
fixed point corresponding to a continuous phase transition exists
only for $n>365.9$, which suggests that for physical parameter
values the transition is first order even in the type-II case. They
corroborated this conclusion by performing a large-$n$ expansion for
fixed $d=3$. To first order in $1/n$, the critical exponent $\nu$ is
positive only for $n>9.72$, which again strongly suggests that the
transition in the physical case $n=2$ is first order. For
superconductors, due to the large correlation length and the
correspondingly small size of the fluctuations, the size of the
first order discontinuity is too small to be observable. For the
analogous smectic-A to nematic transition in liquid crystals, on the
other hand, it was predicted to be much larger, and, indeed,
experimentally observable. In contrast to this theoretical
prediction, experiments on liquid crystals showed, and continue to
show to this day, a clear second order
transition.\cite{Garland_Nounesis_1993} This prompted a
re-examination of the theory by Dasgupta and
Halperin.\cite{Dasgupta_Halperin_1981} Using Monte Carlo data and
duality arguments, these authors argued that a strongly type-II
superconductor in $d=3$ should show a second order transition after
all. A Monte-Carlo study of the intermediate region suggested that
the transition is first order in the strongly type-I region and
continuous in the strongly type-II region, with a tricritical point
separating the two regimes.\cite{Bartholomew_1983} Why the
continuous transition does not show as a critical fixed point in a
$4-\epsilon$ expansion is not quite clear. One possible explanation
is that the fixed point is not perturbatively accessible. Another is
that the critical value of $n$ in the generalized model, which is
close to 366 near $d=4$ where the $\epsilon$-expansion is
controlled, decreases to a value less than 2 as the dimension is
lowered to the physical value $d=3$. Herbut and
Tesanovic\cite{Herbut_Tesanovica, Herbut_Tesanovicb} have shown that
the critical value of $n$, above which there is a second order
transition, decreases rapidly with increasing $\epsilon$, and that a
RG analysis in a fixed dimension $d=3$ to one-loop order does yield
a critical fixed point for systems that are sufficiently strongly
type II.

Recently there has been substantial interest in unconventional
superconductivity. In particular, Sr$_2$RuO$_4$ has emerged as a convincing
case of $p\,$-wave superconductivity,\cite{Nelson_et_al_2004, Rice_2004} and
UGe$_2$ is another candidate.\cite{Machida_Ohmi_2001} This raises the question
whether for such systems the fluctuation-induced first order mechanism is also
applicable, or whether the additional order parameter degrees of freedom allow
for a critical RG fixed point, signalizing a second order transition, even
though no such fixed point is found in the $s$-wave case. Here we investigate
this problem. By conducting an analysis for $p\,$-wave superconductors
analogous to the one of Ref.\ \onlinecite{Halperin_Lubensky_Ma_1974} we find
that there is no critical fixed point, as in the $s$-wave case. This analysis
thus also suggests a first order transition, as in the $s$-wave case, although
the restrictions are somewhat less stringent. Presumably, the same reservations
regarding non-perturbative fixed points that are suspected to be relevant for
the $s$-wave case apply here as well.

This paper is organized as follows. In Sec.\ \ref{sec:II} we define our model
and derive the mean-field phase diagram. In Sec.\ \ref{sec:III} we determine
the nature of the phase transition. We do so first in a renormalized mean-field
approximation that neglects fluctuations of the superconducting order
parameter. We then take such fluctuations into account, first by means of a RG
analysis in $d=4-\epsilon$ dimensions, and then by means of a $1/n$-expansion.
In Sec.\ \ref{sec:IV} we summarize our results.

\section{Model}
\label{sec:II}

Let us consider a Landau-Ginzburg-Wilson (LGW) functional appropriate for
describing spin-triplet superconducting order. The superconducting order
parameter is conveniently written as a matrix in spin space,
\cite{Vollhardt_Woelfle_1990} $\Delta_{\alpha\beta}({\bm k}) = \sum_{\mu=1}^{3}
d_{\mu}({\bm k})\left(\sigma_{\mu} i\sigma_2\right)_{\alpha\beta}$. Here
$\sigma_{1,2,3}$ are the Pauli matrices, ${\bm k}$ is a wave vector, and the
$d_{\mu}$ are the components of a complex $3$-vector ${\bm d}({\bm k})$.
$p\,$-wave symmetry implies $d_{\mu}({\bm k}) = \sum_{j=1}^{3} d_{\mu j}{\hat
k}_j$, with $\hat{\bm k}$ a unit wave vector. The tensor field $d_{\mu j}({\bm
x})$ is the general order parameter for a spin-triplet $p\,$-wave
superconductor and it allows for a very rich phenomenology. For definiteness we
will constrain our discussion to a simplified order parameter describing the
so-called $\beta$-state,\cite{Vollhardt_Woelfle_1990} which has been proposed
to be an appropriate description of UGe$_2$.\cite{Machida_Ohmi_2001} It is
given by a tensor product $d = {\bm\psi}\otimes{\bm\phi}$ of a complex vector
${\bm\psi}$ in spin space and a real unit vector ${\bm\phi}$ in orbital space.
The ground state is given by ${\bm\psi} = \Delta_0(1,i,0)$, ${\bm\phi} =
(0,0,1)$. In a weak-coupling approximation that neglects terms of higher than
bilinear order in $\psi^2$, $\phi^2$, and $\nabla^2$ the action depends only on
${\bm\psi}$,
\bea
S &=& \int d{\bm x}\ \Bigl[ t\vert{\bm\psi}\vert^2 + c\vert{\bm
D}{\bm\psi}\vert^2 + u\vert{\bm\psi}\vert^4 +
v\vert{\bm\psi}\times{\bm\psi}^*\vert^2
\nonumber\\
&& \hskip 50pt + \frac{1}{8\pi\mu}({\bm\nabla}\times{\bm A})^2 \Bigr].
\label{eq:2.1}
\eea
Here ${\bm A}$ is the vector potential, ${\bm D} = {\bm\nabla} - iq{\bm A}$ is
the gauge invariant gradient with $q = 2e$ the Cooper pair charge (we use units
such that Planck's constant and the speed of light are unity), and $\vert{\bm
D}{\bm\psi}\vert^2 = (D_i\psi_{\alpha})(D_i^*\psi_{\alpha}^*)$ with summations
over $i$ and $\alpha$ implied. $\mu$ is the normal-state magnetic permeability,
and $t$, $c$, $u$, and $v$ are the parameters of the LGW functional. The fields
${\bm\psi}$ and ${\bm A}$ are understood to be functions of the position ${\bm
x}$.

For later reference we now generalize the vector ${\bm\psi}$ from a complex
$3$-vector to a complex $m$-vector with components $\psi_{\alpha}$, so that the
total number of order-parameter degrees of freedom is $n=2m$. In order to
generalize the term with coupling constant $v$ we use of the following identity
for $3$-vectors,
\be
\vert{\bm\psi}\times{\bm\psi}^*\vert^2 = \psi_{\alpha}\psi^*_{\alpha}
\psi_{\beta}\psi^*_{\beta} - \psi_{\alpha}\psi_{\alpha}
\psi^*_{\beta}\psi^*_{\beta}
\label{eq:2.2}
\ee
and notice that the right-hand side is well defined for a complex $m$-vector.
Our generalized action now reads
\bea
S &=& \int d{\bm x}\ \Bigl[t\,\psi_{\alpha}\psi^*_{\alpha} +
c\,(D_i\psi_{\alpha})(D_i^*\psi^*_{\alpha}) -
v\,\psi_{\alpha}\psi_{\alpha}\psi^*_{\beta}\psi^*_{\beta}
\nonumber\\
&& \hskip -20pt + (u+v)\,\psi_{\alpha}\psi^*_{\alpha}\psi_{\beta}\psi^*_{\beta}
+ \frac{1}{8\pi\mu}\,\epsilon_{ijk}(\partial_j A_k) \epsilon_{ilm}(\partial_l
A_m) \Bigr]\ , \nonumber\\
\label{eq:2.3}
\eea
with $\alpha,\beta = 1,\ldots m$; $i,j,\ldots = 1,2,3$; and summation over
repeated indices implied. In addition to the generalization of the order
parameter to an $m$-vector we will also consider the system in a spatial
dimension $d$ close to $d=4$. The physical case of interest is $m=d=3$.

\section{Nature of the phase transition}
\label{sec:III}

\subsection{Mean-field approximation}
\label{subsec:III.A}

The simplest possible approximation ignores both the fluctuations of the order
parameter field ${\bm\psi}$ and the electromagnetic fluctuations described by
the vector potential ${\bm A}$. The order parameter is then a constant,
${\bm\psi}({\bm x}) \equiv {\bm\psi}$, and the free energy density $f$ reduces
to
\be
f = t\,\vert{\bm\psi}\vert^2 + u\,\vert{\bm\psi}\vert^4 +
v\,\vert{\bm\psi}\times{\bm\psi}^*\vert^2\ .
\label{eq:3.1}
\ee

In order to determine the phase diagram we parameterize the order parameter as
follows,\cite{Knigavko_Rosenstein_Chen_1999}
\be
{\bm\psi} = \psi_0\,\left({\hat n}\cos\phi + i\,{\hat m}\sin\phi\right).
\label{eq:3.2}
\ee
Here $\psi_0$ is a real-valued amplitude, ${\hat n}$ and ${\hat m}$ are
independent real unit vectors, and $\phi$ is a phase angle. The free energy
density can then be written
\be
f = t\,\psi_0^2 + u\,\psi_0^4 + v\,\psi_0^4\,\left(1 - ({\hat n}\cdot{\hat
m})^2\right)\,\sin^2 2\phi\ .
\label{eq:3.3}
\ee

We now need to distinguish between two cases.

\smallskip
{\it Case 1: $v > 0$.} The free energy is minimized by ${\hat n} = {\hat m}$,
and $\psi_0^2 = -t/2u$. The condition $u > 0$ must be fulfilled for the system
to be stable.

\smallskip
{\it Case 2: $v < 0$.} The free energy is minimized by ${\hat n} \perp {\hat
m}$ and $\phi = \pi/4$, and $\psi_0 = -t/2(u + v)$. The condition $u + v
> 0$ must be satisfied for the system to be stable.

\begin{figure}[t,b,h]
\includegraphics[width=5.0cm]{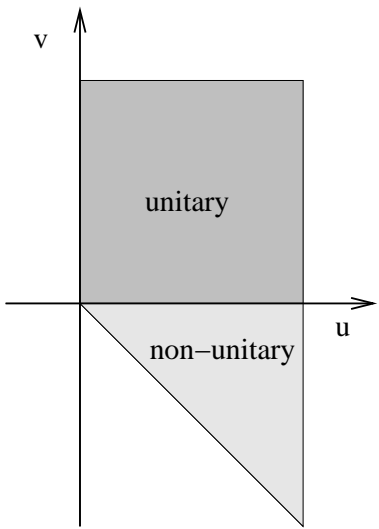}
\caption{Mean-field phase diagram of a $p\,$-wave superconductor as described
 by Eq.\ (\ref{eq:2.1}). See the text for additional information.}
\label{fig:1}
\end{figure}

\smallskip
The first case implies ${\bm\psi}\times{\bm\psi}^* = 0$. This is referred to as
the unitary phase. In the second case, ${\bm\psi}\times{\bm\psi}^* \neq 0$,
which is referred to as the non-unitary phase. In either case, mean-field
theory predicts a continuous phase transition from the disordered phase to an
ordered phase at $t=0$. The mean-field phase diagram in the $u$-$v$ plane is
shown in Fig.\ \ref{fig:1}.

\subsection{Renormalized mean-field theory}
\label{subsec:III.B}

A better approximation is to still treat the order parameter as a constant,
${\bm\psi}({\bm x}) \equiv {\bm\psi}$, but to keep the electromagnetic
fluctuations. The part of the action that depends on the vector potential then
takes the form
\bse
\label{eqs:3.4}
\be
S_A = \frac{1}{8\pi\mu} \int d{\bm x}\ \left[k_{\lambda}^2 {\bm A}^2({\bm x}) +
\left({\bm\nabla}\times{\bm A}({\bm x})\right)\right],
\label{eq:3.4a}
\ee
where
\be
k_{\lambda} = (8\pi\mu c q^2\vert{\bm\psi}\vert^2)^{1/2}
\label{eq:3.4b}
\ee
\ese
is the inverse London penetration depth. Since ${\bm A}$ enters $S_A$ only
quadratically, it can be integrated out exactly,\cite{gauge_fixing_footnote}
and the technical development is identical to the $s$-wave
case.\cite{Halperin_Lubensky_Ma_1974, Chen_et_al_1978} The result for the
leading terms in powers of $\vert{\bm\psi}\vert^2$ in $d=3$ is
\be
f = t\,\vert{\bm\psi}\vert^2 + u\,\vert{\bm\psi}\vert^4 -
w\,(\vert{\bm\psi}\vert^2)^{3/2} + v\,\vert{\bm\psi}\times{\bm\psi}^*\vert^2\ .
\label{eq:3.5}
\ee
Here $w \propto \sqrt{\mu q^2}$ is a positive coupling constant whose presence
drives the transition into either of the ordered phases first order.

There are several interesting aspects of this result. First, the additional
term in the mean-field free energy, with coupling constant $w$, is not analytic
in $\vert{\bm\psi}\vert^2$. This is a result of integrating out the vector
potential, which is a soft or massless fluctuation. Second, the resulting
first-order transition is an example of what is known as the Coleman-Weinberg
mechanism in particle physics,\cite{Coleman_Weinberg_1973} or a
fluctuation-induced first-order transition in statistical
mechanics.\cite{Halperin_Lubensky_Ma_1974}

Let us discuss the validity of the renormalized mean-field theory. The length
scale given by the London penetration depth $\lambda = k_{\lambda}^{-1}$ needs
to be compared with the second length scale that characterizes the action, Eq.\
(\ref{eq:2.1}), which is the superconducting coherence length $\xi =
\sqrt{c/\vert t\vert}$. The ratio $\kappa = \lambda/\xi$ is the Landau-Ginzburg
parameter. For $\kappa \to 0$, order parameter fluctuations are negligible
(this is the limit of an extreme type-I superconductor), and the renormalized
mean-field theory become exact. For nonzero values of $\kappa$ the fluctuations
of the order parameter cannot be neglected, and the question arises whether or
not they change the first-order nature of the transition. We will investigate
this question next by means of two different technical approaches.

\subsection{$\epsilon$-expansion about $d=4$}
\label{subsec:III.C}

\begin{figure}[t,b,h]
\includegraphics[width=8.0cm]{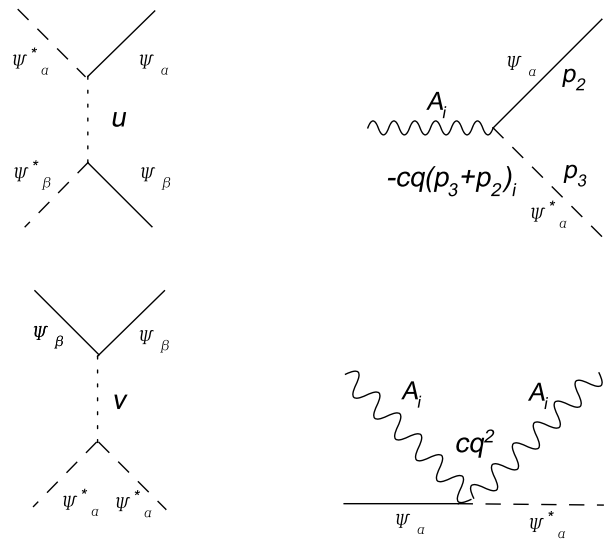}
\caption{Interaction vertices as given by the action, Eq.\ (\ref{eq:2.3}).
Solid and dashed lines denote the electron fields ${\bm\psi}$ and
${\bm\psi}^*$, respectively, and wavy lines denote the gauge field ${\bm A}$.
The dotted lines serve to separate different components of the electron
fields.}
\label{fig:2}
\end{figure}

\begin{figure}[t,b,h]
\includegraphics[width=9.0cm]{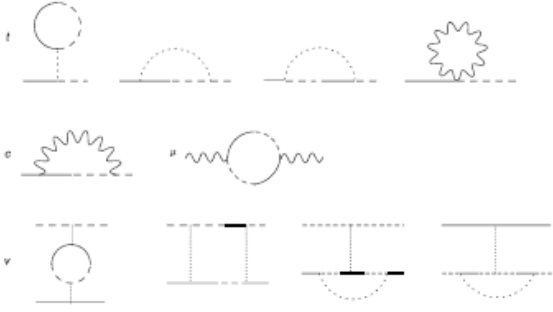}
\caption{One-loop diagrams that renormalize the coupling constants $t$, $c$,
 $u$, $v$, $q$, and $\mu$ in Eq.\ (\ref{eq:2.3}).}
\label{fig:3}
\end{figure}

We first perform a momentum-shell renormalization-group (RG) analysis of the
action, Eq.\ (\ref{eq:2.3}), in $d=4-\epsilon$ dimensions. The propagators can
be read off the action, Eq.\ (\ref{eq:2.3}). For the $\psi$-propagator we have
\bse
\label{eqs:3.6}
\be
\langle \psi_{\alpha}({\bm k})\,\psi^*_{\beta}(-{\bm k})\rangle =
\frac{\delta_{\alpha\beta}}{t + ck^2}\ .
\label{eq:3.6a}
\ee
In Coulomb gauge, ${\bm\nabla}\cdot{\bm A} = 0$, one finds for the gauge field
propagator
\be
\langle A_i({\bm k})\,A_j(-{\bm k})\rangle = 4\pi\mu\, \frac{\delta_{ij} -
{\hat k}_i {\hat k}_j}{k^2}\ ,
\label{eq:3.6b}
\ee
\ese
where ${\hat k}_i$ denotes the components of the unit vector ${\hat{\bm k}} =
{\bm k}/k$. $\langle \ldots \rangle$ denotes an average with respect to the
Gaussian part of the action, Eq.\ (\ref{eq:2.3}). The vertices as given by the
action, Eq.\ (\ref{eq:2.3}), are shown graphically in Fig.\ \ref{fig:2},
and the one-loop diagrams that renormalize the various coupling constants in
the action are shown in Fig.\ \ref{fig:3}.
The calculation is now a straightforward generalization of the one for the
s-wave case, Ref.\ \onlinecite{Halperin_Lubensky_Ma_1974}. We define the scale
dimension of a length $L$ to be $[L] = -1$, and exponents $\eta$ and $\eta_A$
by choosing the scale dimensions of the fields ${\bm\psi}$ and ${\bm A}$ to be
$[{\bm\psi}({\bm x})] = (d-2+\eta)/2$ and $[{\bm A}({\bm x})] =
(d-2+\eta_A)/2$, respectively. We find RG recursion relations
\bse
\label{eqs:3.7}
\bea
\frac{dt}{dl} &=& (2-\eta)\,t + 3\,c\,q^2\mu + \frac{(n+2)u + 4v}{t+c}\ ,
 \label{eq:3.7a}\\
\frac{du}{dl} &=& (\epsilon-2\eta)\,u - \frac{(n+8)u^2 + 8uv + 8v^2}{(t+c)^2} -
3\,c^{2}q^{4}\mu^{2}, \nonumber\\
\label{eq:3.7b}\\
\frac{dv}{dl} &=& (\epsilon-2\eta)\,v - \frac{nv^2 + 12uv}{(t+c)^2}\ ,
\label{eq:3.7c}\\
\frac{dc}{dl} &=& -\eta\,c - 3\frac{c^2 q^2 \mu}{t+c}\ ,
\label{eq:3.7d}\\
\frac{d\mu}{dl} &=& \eta_A\,\mu - \frac{n}{6}\, \frac{(3t+c)c^3 q^2
\mu^2}{(t+c)^4}\ ,
\label{eq:3.7e}\\
\frac{dq}{dl} &=& \frac{1}{2}\,(\epsilon - \eta_A)\,q.
\label{eq:3.7f}
\eea
\ese
Here $l = \ln b$ with $b$ the length rescaling parameter, we have redefined
$4\pi\mu \to \mu$, and we have absorbed a common geometric factor in the
coupling constants $u$, $v$, and $\mu$. For $v=0$, these flow equations reduce
to those of Ref.\ \onlinecite{Halperin_Lubensky_Ma_1974}, as they should.

We now look for fixed points (FPs) of the Eqs.\ (\ref{eqs:3.7}). Since we are
interested in superconductors (as opposed to superfluids), we are looking for a
FP where the charge is nonzero. Equation\ (\ref{eq:3.7f}) immediately yields
\bse
\label{eqs:3.8}
\be
\eta_A = \epsilon.
\label{eq:3.8a}
\ee
Anticipating a FP value $t^*$ of $t$ that is of $O(\epsilon)$, Eq.\
(\ref{eq:3.7e}) then implies $(q^2\mu)^* = 6\epsilon/n + O(\epsilon^2)$. If we
choose $\eta$ such that $c$ is not renormalized, Eq.\ (\ref{eq:3.7d}) in turn
yields
\be
\eta = -18\epsilon/n + O(\epsilon^2).
\label{eq:3.8b}
\ee
\ese
We now look for FP values of $u$ and $v$ that are of $O(\epsilon)$. Equation
(\ref{eq:3.7a}) then indeed yields $t^* = O(\epsilon)$, so the remaining task
is to consider Eqs.\ (\ref{eq:3.7b}, \ref{eq:3.7c}). Let us define $x =
u^*/\epsilon c^2$ and $y = v^*/\epsilon c^2$. The remaining FP equations then
read
\bse
\label{eqs:3.9}
\bea
(n+8)x^2 + 8xy + 8y^2 - \left(1+\frac{36}{n}\right)x + \frac{108}{n^2} &=& 0,
\nonumber\\
\label{eq:3.9a}\\
ny^2 + 12xy - \left(1+\frac{36}{n}\right)y &=& 0. \nonumber\\
\label{eq:3.9b}
\eea
\ese

This set of two coupled quadratic equations has four solutions. Two of these
are given by
\bse
\label{eqs:3.10}
\be
y = 0,
\label{eq:3.10a}
\ee
and $x$ a solution of
\be
(n+8)x^2 - \left(1 + \frac{36}{n}\right)x + \frac{108}{n^2} = 0.
\label{eq:3.10b}
\ee
\ese
Equation (\ref{eq:3.10b}) is the same condition as in the s-wave case, Ref. \
\onlinecite{Halperin_Lubensky_Ma_1974}. It has a real positive solution for $n
> n_{\text c}^{\text s} \approx 365.9$. In this case, $v_0^*=0$, and $u_0^* = c^2
x\epsilon + O(\epsilon^2) > 0$. We will refer to these as the s-wave FPs. The
only other real solutions of Eq.\ (\ref{eq:3.10b}) occur in the unphysical
region $n \alt -5.9$.

For $y\neq 0$, we have
\bse
\label{eqs:3.11}
\be
x = \frac{1}{12}\left(1 + \frac{36}{n} - ny\right),
\label{eq:3.11a}
\ee
and $y$ a solution of
\bea
&&\hskip -20pt n^2(n^3 + 8n^2 - 96n + 1152)y^2
\nonumber\\
&& \hskip 0pt - 2n(n^3 + 38n^2 + 24n - 1728)y
\nonumber\\
&& \hskip 0pt + (n^3 + 68n^2 + 1008 n + 10368) = 0.
\label{eq:3.11b}
\eea
\ese
For positive values of $n$, this equation has real solutions only for $n >
n_{\text c}^{\text p} \approx 420.9$, which leads to two FPs that we refer to
as the p-wave FPs. A stability analysis shows that of the four FPs found, the
only stable one is the p-wave FP with the larger value of $y$. Linearizing
about his FP yields the critical exponent $\nu$ for the corresponding
continuous phase transition. For general $n$, the expression is very
complicated. In the limit of large $n$, one finds
\be
1/\nu = 2 - (1 - 1/n + O(1/n^2))\epsilon + O(\epsilon^2).
\label{eq:3.12}
\ee

We conclude that for large $n$, and close to $d=4$, the phase transition in
p-wave superconductors is continuous, as it is in the s-wave case, and that the
p-wave case is in a different universality class. The analysis also suggests
that for physical parameter values, $n=6$ and $d=3$, the transition is unlikely
to be continuous. Of course, the same caveats as in the s-wave case apply with
respect to the interpretation of this result. As was shown in Ref.\
\onlinecite{Halperin_Lubensky_Ma_1974}, additional information can be obtained
by means of an expansion in $1/n$ in $d=3$, and we perform such an analysis in
the next subsection.

\subsection{$1/n$-expansion in $d=3$}
\label{subsec:III.D}

The technique of the $1/n$ expansion in a fixed dimension was developed by
Ma\cite{Ma_1973} for an neutral $n$-component vector field, and it was
generalized to the presence of a gauge field in Ref.\
\onlinecite{Halperin_Lubensky_Ma_1974}. The basic idea is as follows. Consider
the fully dressed or renormalized counterpart $G$ of the Gaussian propagator
given in Eq.\ (\ref{eq:3.6a}). Its inverse can be written in terms of a self
energy $\Sigma$,
\be
G^{-1}(t,k) = t + ck^2 + \Sigma(t,k).
\label{eq:3.13}
\ee
At zero wave number, we have
\bse
\label{eqs:3.14}
\be
G^{-1}(t,0) = t + \Sigma(t,0) \equiv t_{\text{R}},
\label{eq:3.14a}
\ee
where $t_{\text{R}}$ is the renormalized counterpart of $t$. For $t$
approaching its critical value $t_{\text{c}}$, it vanishes according to a power
law
\be
t_{\text{R}} \propto (t-t_{\text{c}})^{\gamma},
\label{eq:3.14b}
\ee
\ese
characterized by the critical exponent $\gamma$. This implies $t_{\text{c}} =
-\Sigma(t=t_{\text{c}},0) = -\Sigma(t_{\text{R}}=0,0)$, which allows to
determine $t_{\text{c}}$ order by order in some perturbative scheme. At
criticality, Eq.\ (\ref{eq:3.13}) can thus be rewritten
\bea
G^{-1}(t=t_{\text{c}},k) &=& ck^2 + \Sigma(t=t_{\text{c}},k) -
\Sigma(t=t_{\text{c}},0)
\nonumber\\
&\propto& k^{2-\eta} = k^2\left[1-\eta\ln k + O(\eta^2)\right], \nonumber\\
\label{eq:3.15}
\eea
with the critical exponent $\eta$. Now consider a perturbative expansion for
$\Sigma$ with $1/n$ as the small parameter. Assuming that $\eta$ is small, it
thus can be determined perturbatively from the wave number dependence of
$\Sigma(t=t_{\text{c}},k) - \Sigma(t=t_{\text{c}},0)$. To zeroth order in this
expansion, there is no contribution, so $G^{-1}(t=t_{\text{c}},k) \propto k^2$,
and $\eta = 0$. To first order in $1/n$, there is a contribution, as one would
expect from the result of the $\epsilon$ expansion, Eq.\ (\ref{eq:3.8b}).

Similarly, $\gamma$ can be obtained perturbatively from the behavior of the
self energy at $k=0$. From Eqs.\ (\ref{eq:3.13}, \ref{eqs:3.14}) we have
\be
t_{\text{R}} + \Sigma(t_{\text{R}}=0,0) - \Sigma(t_{\text{R}},0) = t -
t_{\text{c}} \propto t_{\text{R}}^{1/\gamma}.
\label{eq:3.16}
\ee
To zeroth order in $1/n$ one finds, in $d=3$, $\Sigma(t_{\text{R}}=0,0) -
\Sigma(t_{\text{R}},0) \propto t_{\text{R}}^{1/2}$. That is, $\gamma = 2$,
which is result for the spherical model. If we write corrections to this result
in terms of an exponent $\zeta$, $1/\gamma = 1/2 - \zeta$, we have
\be
\Sigma(t_{\text{R}}=0,0) - \Sigma(t_{\text{R}},0) \propto
t_{\text{R}}^{1/2-\zeta} = t_{\text{R}}^{1/2}[1 - \zeta\ln t_{\text{R}} +
O(\zeta^2)].
\label{eq:3.17}
\ee
$\zeta$ can just be obtained as the prefactor of a dependence of the self
energy on $t_{\text{R}}^{1/2}\ln t_{\text{R}}$.

For s-wave superconductors, this calculation has been performed in
Ref.\ \onlinecite{Halperin_Lubensky_Ma_1974}. For the current
problem, the calculation is straightforward.
This calculation yields \bse \label{eqs:3.18} \bea \eta &=&
\frac{-104}{3\pi^2}\,\frac{1}{n} + O(1/n^2),
\label{eq:3.18a}\\
\gamma &=& 2\left(1 - \frac{100}{\pi^2}\,\frac{1}{n} + O(1/n^2)\right).
\label{eq:3.18b}
\eea
All other static critical exponents can be obtained from these two by means of
scaling relations. In particular, for the correlation length exponent $\nu$ we
have
\be
\nu = 1 - \frac{352}{3\pi^2}\,\frac{1}{n} + O(1/n^2).
\label{eq:3.18c}
\ee
\ese
To this order, $\nu$ is positive for $n > 352/3\pi^2 \approx 11.9$, which
suggests again that the transition is first order for the physical case $n=6$
in $d=3$.

\section{Summary}
\label{sec:IV}

In summary, we have considered the critical behavior of a p-wave superconductor
in both an $\epsilon$ expansion about $d=4$, and in a $1/n$ expansion in $d=3$,
in analogy to the analysis of s-wave superconductors in Ref.\
\onlinecite{Halperin_Lubensky_Ma_1974}. We have found that the results are
qualitatively the same: both methods suggest that, for physical parameter
values, the superconducting transition is fluctuation-induced first order in
nature. For a model with an $n$-component order parameter, the suppression of
fluctuations for sufficiently large $n$ leads to a continuous transition that
is in a different universality class than the corresponding transition in the
s-wave case. To first order in $\epsilon = 4-d$, the critical value of $n$ that
separates the first-order and second-order cases is $n_{\text{c,p}} = 420.9$
for the p-wave case, whereas in the s-wave case one has $n_{\text{c,s}} =
365.9$. In a $1/n$ expansion in $d=3$, the critical $n$-value in the p-wave
case is equal to $11.9$, compared to $9.72$ in the s-wave case. These values
have to be compared to the physical values $n=2$ and $n=6$ in the s-wave and
p-wave cases, respectively. The caveats related to critical fixed points that
are not accessible by either perturbative method that have been discussed for
the s-wave case\cite{Dasgupta_Halperin_1981, Bartholomew_1983} apply to the
p-wave case as well.

\section{acknowledgments}

This work was supported by the NSF under grant No. DMR-05-29966.

\bibliography{p_wave}

\end{document}